\documentclass[5p]{elsarticle}
\pdfoutput=1
\usepackage{graphicx}
\usepackage{amsmath} 
\usepackage{amsfonts}
\usepackage{amssymb}
\usepackage{hyperref}
\usepackage{subfigure}
\usepackage{longtable,lscape}
\usepackage[normalem]{ulem} 
\usepackage{hepunits}
\usepackage{multirow} 
\biboptions{sort&compress}

\newcommand{\Trace}{\operatorname{Tr}}

\newcommand{\BigO}{\operatorname{O}}
\newcommand{\etal}{\emph{et al.}}

\newcommand{\eq}{Eq.~}
\newcommand{\fig}{Fig.~}
\newcommand{\tab}{Tab.~}
\newcommand{\fm}{{\rm \,fm}}

\begin{document}

\title{Topology of dynamical lattice configurations\\
including results from dynamical overlap fermions}
\author[a]{Falk Bruckmann}
\author[a]{Florian Gruber\corref{cor1}}
\ead{florian.gruber@physik.uni-regensburg.de}
\author[a,b]{Nigel Cundy}
\author[a]{Andreas Sch\"afer}
\author[c]{Thomas Lippert}
\address[a]{Institut f\"ur Theoretische Physik, Universit\"at Regensburg, D-93040 Regensburg, Germany}
\address[b]{Department of Physics and Astronomy, Seoul National University, Seoul, 151-747, South Korea}
\address[c]{J\"ulich Supercomputing Centre, Forschungszentrum J\"ulich GmbH, D-52425 J\"ulich, Germany}
\cortext[cor1]{Corresponding author}
\date{\today}

\begin{abstract}
We investigate how the topological charge density in lattice QCD simulations is affected by violations of chiral symmetry caused by the fermion action. To this end we compare lattice configurations generated with a number of different actions including first configurations generated with exact dynamical overlap quarks. We visualize the topological profiles after mild smearing. In the topological charge correlator we measure the size of the positive core, which is known to shrink to zero extension in the continuum limit. To leading order we find the core size to scale linearly with the lattice spacing with the same coefficient for all actions, even including quenched simulations. In the subleading term the different actions vary over a range of about 10\%. Our findings suggest that non-chiral lattice actions at current lattice spacings do not differ much for observables related to topology, both among themselves and compared to overlap fermions.
\end{abstract}

\maketitle

\section{Introduction}
\sloppy
The topological charge density has always been an important observable to characterize  the vacuum of Quantum Chromodynamics (QCD). 
Instantons are the most natural candidates carrying topological charge. Models based on them are intuitive, e.g.\ induce a chiral 
condensate through near zero modes, but also limited, since it became clear that the instanton ensemble cannot be dilute in reality. Such models are also not complete, as instantons do not explain confinement. 

In recent years a more elaborated view has emerged. The relevant topological degrees of freedom depend on the scale. While instantons are seen in the IR (after strong filtering), lower dimensional objects are present in the UV, for a review see \cite{Forcrand2007}. A laminar structure of them seems needed to match the known behavior of the topological correlator \cite{Horvath:2003yj} (see below). This picture is based in particular on the analysis of lattice configurations.

Lattice QCD has seen progress towards smaller lattice spacings and realistic pion masses. Chiral symmetry, however, is realized very differently in these simulations. Most of the fermion actions used keep only an approximation of it. The reason is that the only known option to implement chiral symmetry exactly are Ginsparg-Wilson fermions, usually realized as overlap fermions, which are numerically very expensive.

Since chiral symmetry is a key feature of QCD, it is not clear a priori, how harmful this approximation is for physics results. However, the fundamental aim of lattice QCD is to avoid uncontrolled approximations and, therefore, it is necessary to perform benchmark comparisons with dynamical overlap fermions to quantify the size of systematic errors due to chiral symmetry violation.

In this Letter we analyze the topological charge in simulations obtained with different state-of-the-art lattice fer\-mion actions. 
Topology is intimately connected to chiral symmetry, both through the $U(1)$ anomaly giving mass to the $\eta'$ and through the index theorems linking fermionic zero modes to topological charge. We, therefore, believe that the topological charge density, although it is a purely gluonic quantity, is well suited to test the effects due to violation of chiral symmetry for different fermion actions.
The distance at which the topological charge correlator changes sign seems to be especially robust against small changes in the simulation parameters.

We use the field-theoretic definition of the lattice topological charge 
and compare  topological structures obtained after a few sweeps of smearing (alternative filtering methods, for instance based on chiral lattice Dirac operators, have been shown to largely agree on topological structures and to be sensitive of the difference between quenched and dynamical configurations, see below). We visualize the topological profiles after mild smearing though results for different lattice spacing $a$ are difficult to compare. 
To reach more quantitative results, we compare the size of the positive core in the topological charge correlator and the amplitude of the contact term.

For these quantities we compare the results with those of configurations generated with overlap fer\-mions. These latter simulations are too expensive for comprehensive simulations, but they allow for a benchmark test of the described nature on small lattices. 

Our main conclusions are that the differences between the actions are not very large, even not for quenched simulations. The sensitivity to the lattice spacing $a$ far exceeds that to 
the fermion action used. Therefore, observed differences, e.g. between quenched and dynamical fermion results 
should not be over interpreted. Obviously similar analyses of other quantities are needed 
to decide whether these conclusions are valid in general. 
 
\section{The topological charge density}
\subsection{\ldots in the continuum }
The topological charge and its (Euclidean) density are defined as
\begin{equation}
\label{eq:qdens}
Q=\int d^4 x\, q(x),\quad
q(x)=\frac{1}{32\pi^2}\, \epsilon_{\mu\nu\rho\sigma}{\rm Tr}\Big(F_{\mu\nu}(x)F_{\rho\sigma}(x)\Big).
\end{equation}
In this Letter we focus on the two-point function of the topological charge density
\begin{equation}\label{eq:qdenscorr}
 C_q(x-y)\equiv \langle q(x)q(y)\rangle,
\end{equation}
which integrates to the topological susceptibility
\begin{equation}\label{eq:topsuceptibility}
\chi_{top}=\frac{\langle Q^{2}\rangle}{\text{V}}=\int \langle q(0)q(x)\rangle d^{4}x.
\end{equation}

In the continuum, this correlator is negative for any finite distance $|x-y|>0$. This follows from reflexion-positivity of the theory and the pseudo-scalar nature of the topological charge density \cite{Seiler}. At zero distance, however, we obviously have a positive correlator, as $C_q(0)=\langle q(0)^2\rangle$ is nothing but the mean-square topological charge density.  

This property on its own would be strange enough, but the topological susceptibility is non-negative and vanishes in the chiral limit \cite{Seiler}. Therefore, the positive contact term has to (over)com\-pen\-sa\-te the negative contribution of the correlator in the spacetime integral in \eq(\ref{eq:topsuceptibility}).

The topology of the gauge field nicely reflects itself in fermionic observables. Due to the famous index theorems, the total topological charge gives the difference of numbers of zero modes with left-handed and right-handed chirality, $Q=n_L-n_R$, of the chiral (massless) Dirac operator.

\subsection{\ldots\ and on the lattice}
The topological charge, as a genuine continuum quantity, is not defined in a strict sense
on a lattice. Of course, the expectation is that its properties will be recovered in the 
continuum limit. At finite lattice spacing, however, it cannot be defined without ambiguities.

Throughout this Letter we will use the discretized topological charge density of \eq(\ref{eq:qdens}) 
(also called field-theoretic charge density), where $F_{\mu\nu}$ is an improved lattice field strength 
tensor \cite{Bilson-Thompson2003}, which combines $1\times1$, $2\times2$ and $3\times3$ loops to 
achieve $O(a^4)$-improvement at tree-level\footnote{For details on the improvement coefficients see 
\cite{Bilson-Thompson2003}}. This definition is better behaved in the continuum limit than the naive 
discretization with only $1\times1$ loops and gives topological charges $Q=\sum_x q(x)$ closer to integers.

Calculating the topological charge density on a typical Monte Carlo configuration one finds that the 
signal is dominated by short range fluctuations. Thus, an operator which couples strongly to these fluctuations 
is ill-defined. Small variations of the gauge links would lead to big variations of the resulting 
topological density. Hence, filtering methods are necessary to extract the relevant longer range 
degrees of freedom.

Many methods have been developed since the advent of lattice QCD. Amongst them are cooling, APE smearing \cite{Albanese:1987ds}, stout smearing \cite{Morningstar2004a} and variants thereof, as well as techniques based on the low mode truncation of the lattice Laplacian \cite{Bruckmann2005a}.

A different definition of a lattice topology uses the index theorem, which for Ginsparg-Wilson operators is valid\footnote{Still the total topological charge on the lattice is not unique since it depends on the solution of the GW relation, e.g.\ on the kernel of the overlap operator.} and can be formulated in terms of a local density, $q_{\rm ferm}(x)\equiv{\rm Tr}\big( \gamma_5 (D(x,x)/2-1)\big)$ \cite{Niedermayer1999}. This density can be filtered naturally by including in a spectral representation of $D$ only the lowest IR modes \cite{Horvath:2002yn,Ilgenfritz2007a} (see also \cite{Perez:2011tx} for a recent filtering method relying on adjoint fermions).

The structures which emerge after these filtering procedures are not unique. They strongly depend on parameters like numbers of iterations or mode number. However, in the weak filtering regime one finds that different methods reveal similar local structures if one carefully matches their parameters \cite{Bruckmann2007c,Bruckmann2009a,Bruckmann2010a,Ilgenfritz:2008ia}.

\section{Smearing}

In this work we use the improved stout smearing algorithm \cite{Moran2008,Ilgenfritz:2008ia}. This is an iterative procedure acting on the gauge links. Each step consists of a simultaneous update of all links by       
\begin{table*}[t!!!]
\centering
\begin{tabular}{ccccccc}\hline
\textbf{fermion (gauge) action }&$\mathbf{N_{f}}$ & $\mathbf{a[\fm]}$& $\mathbf{V[a^4]}$ & $\mathbf{m_{\pi} [\MeV]}$ &$\mathbf{N_{\rm conf}}$&{\bf Ref.}\\ \hline
twisted mass  (Sym) &2& $0.10-0.063$ & $24^3 48\, - \,32^3 64$ &$\approx 500 $&20 &\cite{Baron:2009wt}\\ 
twisted mass  (Sym) &2& $0.10-0.051$ & $20^348\, -\,32^364$ &$ \approx 280 $&20 &\cite{Baron:2009wt}\\  
np imp. clover (plaq.) &2& $0.11-0.07$ & $24^3 48$ &$\approx 500$&20& \cite{Gockeler2006}\\  
np imp. clover (plaq.) &2& $0.10-0.07$ & $32^3 64$ &$\approx 250 $&20& \cite{Gockeler2006}\\  
asqtad staggered  (LW)&2+1& $0.15-0.09$ & $16^3 48\, - \, 28^{3}96$ &$\approx 500$&5&\cite{Bernard2007a} \\  
chirally imp. (LW) &2& $0.15$ &  $16^3 32$ &$\approx 500$& 20 &\cite{Gattringer2009} \\ 
top. fixed overlap (Iw)&2 & $0.12$ & $16^3 32$&$\approx 500 $&20&\cite{Aoki2008} \\  
overlap (LW) &2+1& $0.13-0.12$ & $12^3 24$ &$\approx 500$& 15 & \cite{Cundy:2009ae} \\ \hline 
\end{tabular}
\caption{Configurations used in this work. (Abbreviations: Sym = Symanzik , plaq. = plaquette, LW = L\"uscher-Weisz, Iw = Iwasaki)}\label{tab:configs}
\end{table*}
\begin{equation}
 \label{eq:Stoutsmearing}
U^{\rm stout}_{\mu}(x)=\exp(i\rho\,\Omega(x))\cdot U_{\mu}(x),
\end{equation}
Here $\rho$ is a parameter which determines the strength of smearing and $\Omega(x)$ is a hermitian traceless matrix built from the weighted sum of all $1\times 1$ and $2\times 1$ Wilson loops including the old link:
\begin{equation}
\begin{split}
\label{eq:Omegaofstoutsmearing}
  \Omega(x)=\sum \Bigg[&\frac{5-2\epsilon}{3}\,W_{1\times 1}(x)\\&+ \frac{1-\epsilon}{12} \bigg(W_{2\times 1}(x) + W_{1\times 2} (x)\bigg)\Bigg]\Bigg|_{{\shortstack{\rm \tiny hermitean, \\[-2 pt] \tiny traceless}}                                                                                                  }\;,
\end{split}
\end{equation}
where we used $\tfrac{i}{2}(\Omega_{\mu}^{\dagger}(x)-\Omega_{\mu}(x))-\tfrac{i}{6}\Trace \lbrace \Omega_{\mu}^{\dagger}(x)-\Omega_{\mu}(x) \rbrace$ as the hermitian traceless projection.

While other smearing methods, as well as cooling, destroy topological structures in the long run, this algorithm is designed to preserve instanton-like objects in a wide range of their size parameter, if we choose $\epsilon =-0.25$ and $\rho=0.06$ \cite{Moran2008}. Moreover, 5 steps of this improved stout smearing were found to produce 
a topological charge density very similar to the fermionic one, $q_{\rm ferm}$ \cite{Ilgenfritz:2008ia}, which is why we stick to these parameters in this work.

\section{Lattice configurations}

\subsection{Overview of lattice configurations used in this work}
We use full dynamical $N_{f}=2$ and $N_{f}=2+1$ flavor configurations from different fermion formulations. They are available through the International Lattice Data Grid \cite{ildg} in a wide range of lattice spacings, lattice volumes and pion masses. 

The fermion actions in these simulations may be classified with respect to their chiral symmetry. Variants of Wilson fermions like twisted mass \cite{Baron:2009wt} and nonperturbative clover \cite{Gockeler2006} break chiral symmetry explicitly (by a lattice artefact). Staggered fermion actions like asqtad \cite{Bernard2007a} possess a remnant chiral symmetry. For those actions we have enough data to perform an extrapolation $a\to 0$. 

We also use chirally improved fermions, which are an approximate solution to the Ginsparg-Wilson relation \cite{Gattringer2009} and as exact solutions we  take overlap fermions with an extra topology fixing term \cite{Aoki2008} and full dynamical overlap fermions, see next subsection. More details on the configurations can be found in \tab\ref{tab:configs}. The pion masses for the configurations used vary between $250$ and 
$500\,\MeV$, which does not seem to be a major problem as little mass dependence is found.  

\subsection{Dynamical overlap configurations}
The overlap Dirac operator \cite{Neuberger:1997fp} at a mass parameter $\mu$ and for some suitable kernel operator $K$ is
\begin{equation}
 D_o = (1+\mu) + (1-\mu) \gamma_5 \text{sign}(K).
\end{equation}
The principle algorithmic challenge in the generation of dynamical overlap gauge field ensembles is the difficulty changing topological charge during the Monte Carlo history. As in the continuum, to move to different topological sectors for a Ginsparg-Wilson lattice gauge theory requires a discontinuous change in the action. For an overlap action, where there is a clear and consistent definition of the topological index, a topological index change is equivalent to a small eigenvalue of the kernel operator crossing zero. From this viewpoint, there are two ways in which topology changes may be suppressed: either the density of low kernel eigenvalues is decreased, or there is a potential barrier between topological sectors. The Hybrid Monte Carlo trajectory can only penetrate this barrier if it has sufficient momentum (there is a clear analogy between the mechanics of the overlap HMC algorithm and the classical mechanics of a particle approaching a finite potential barrier: the only difference is that instead of three spatial dimensions in classical mechanics, the number of `spatial' dimensions is of order of the lattice volume in the molecular dynamics). Changing topology becomes harder for most lattice discretizations as they approach the continuum limit. For overlap fermions we have this challenge at any lattice spacing.  

JLQCD's topology fixed simulations~\cite{Aoki2008} add extra fields to the action which act as a determinant factor $\det\{K^2/(K^2 + \eta^2\}$ in the Boltzmann weight. This has the effect of suppressing (and removing) the low kernel eigenvalues, such that the simulation is carried out in a fixed topological sector, which greatly reduces the computer time required for each trajectory.

The physical effects of the topology fixing are removed in the infinite volume limit~\cite{Aoki2007}, while the suppression of low modes is said to mimic the expected behavior in the continuum limit. Questions remain, however, about the ergodicity of this approach; in particular if the creation of pairs of would be zero modes (for example an instanton/anti-instanton pair) is driven by mixing between low modes of the kernel operator.  

The second possibility is to generate overlap ensembles allowing (and even encouraging as much as the underlying physics will permit) topological charge changes. The standard, and only known, method to modify the Hybrid Monte Carlo algorithm for a discontinuous action is the transmission/reflection algorithm first proposed in~\cite{Fodor:2003bh} and refined in several subsequent papers~\cite{DeGrand:2004nq,Schaefer:2005qg,Cundy:2005pi,Cundy:2005mn,Cundy:2005mr,Cundy:2008zc}, together with the eigenvector differentiation algorithm proposed in~\cite{Cundy:2007df}, which is required to accurately differentiate near degenerate pairs of eigenvectors whose eigenvalues lie close to a discontinuity. The chief focus of current algorithmic research has been to improve the rate of topological tunneling. It can be shown that if the discontinuity of the action at the boundary between topological sectors is $\Delta S$, then, with an initial Gaussian momentum distribution, the probability of tunneling from one topological sector to another scales like $\sim \min(1,e^{-\Delta S})$ for all the possible transmission/reflection algorithms. 

To obtain a high rate of topological charge changes and, presumably, a small topological autocorrelation time, it is crucial to use the form of the action that has the smallest $\Delta S$. When trying to find it one encounters the problem that the pseudo-scalar estimate of the determinant gives a very bad estimate of $\Delta S$~\cite{DeGrand:2004nq}, in particular for small fermion mass ($\Delta S \sim \mu^{-2}$). 
\begin{equation}
\begin{split}
 \langle \Delta& S_{\text{pseudofermion}}\rangle = \big\langle\phi^{\dagger}\big|\frac{1}{D_{\rm o}^{\dagger}D_{\rm o}(\lambda > 0)} - \frac{1}{D_{\rm o}^{\dagger}D_{\rm o}(\lambda < 0)}\big| \phi \big\rangle\\&\quad> \langle \Delta S_{\text{real}}\rangle = 2 \log\big(1-2 (1-\mu)\psi_0^{\dagger}\frac{1}{D_{\rm o}}\psi_0\big),
\end{split}
\end{equation}
where $\psi_0$ is the kernel eigenvector whose eigenvalue chan\-ges sign and $\phi$ is the pseudo\-fermion field used to estimate the determinant.

To improve the pseudofermion approach one can introduce the modified Dirac operator
\begin{equation}
\tilde{D}_{\rm o} = (1+\mu) + (1-\mu) \gamma_5 \text{sign}(K- \Lambda_0),
\end{equation}
for a real parameter $\Lambda_0$ which remains constant over the course of the trajectory. With this operator one can rewrite
\begin{equation}
\det D_{\rm o} = \det\tilde{D}_{\rm o} \det (D_{\rm o}/\tilde{D}_{\rm o}).
\end{equation}
The first term $\det\tilde{D_{\rm o}}$ is continuous during the topological index change and can be estimated using pseudofermions, while the second term $\log \det (D_{\rm o}/\tilde{D_{\rm o}})$ can be calculated without pseudofermions. This factorization leads to a correct estimate for the action discontinuity:
 \begin{equation} \Delta S = \Delta \log \det (D_{\rm o}/\tilde{D_{\rm o}}) = \Delta \log\det D_{\rm o} = \Delta S_{\text{real}}.\end{equation}
 This factorization method (together with other, similar, approaches) thus gives the optimal approach to tunneling through the topological barrier, with the highest transmission rate possible while conserving the molecular dynamics energy. 

Because $\det\tilde{D}_{\rm o}$ at mass zero satisfies a Ginsparg-Wilson relation, and we do not permit the index of $\det\tilde{D}_{\rm o}$ to change during a trajectory, it is possible to run single flavor simulations by using the relation
\begin{equation}
\begin{split}
 \det\tilde{D_{\rm o}} \propto& \det\big(\sqrt{2+2\mu^2} \\&\pm \frac{{1-\mu^2}}{2\sqrt{2+2\mu^2}} (1\pm\gamma_5)\,\text{sign}(K- \Lambda_0) (1\pm \gamma_5)\big),
\end{split}
\end{equation}

where the sign is chosen to avoid the zero modes and the constant of proportionality is a function of the mass and the index of $\det\tilde{D_{\rm o}}$. Because the operator on the right-hand side is hermitian and positive definite, it is possible to use a standard pseudofermion approach to estimate the determinant for a single flavor, without need to employ an expensive rational approximation.

The data used in this study is based on three simulations on $12^324$ lattices using a tadpole improved L\"uscher-Weisz gauge action. Though we can simulate individual quarks we choose two degenerate light quark masses and the strange quark mass is tuned according to the QCDSF's prescription~\cite{Bietenholz:2010jr}. The pion masses, which are restricted by the small lattice volume, were measured to be around $510\,\MeV$, $560\,\MeV$ and $600\,\MeV$, at lattice spacings of around $0.12 \fm$. Over 1000 trajectories were generated for the lightest pion mass, and over 600 trajectories were used for the heavier pion masses. We used a Wilson kernel with one step of improved stout smearing~\cite{Moran2008}. 

The limiting autocorrelation time for these ensembles was, not surprisingly, the topological one. For the lightest mass, there were on average 1.05 attempted topological index changes per trajectory, of which 37.5\% resulted in a change of the topological index. This (naively) corresponds to one topological index change for every 2.8 trajectories of length 0.5. (This excludes trajectories where the topological charge changed twice or where the trajectory was rejected.) The Monte Carlo acceptance rate was around 85\%. Values for the other $12^324$ runs were similar.

\begin{figure}[t!!!]
\centering
\includegraphics[width=0.7\columnwidth,angle=270]{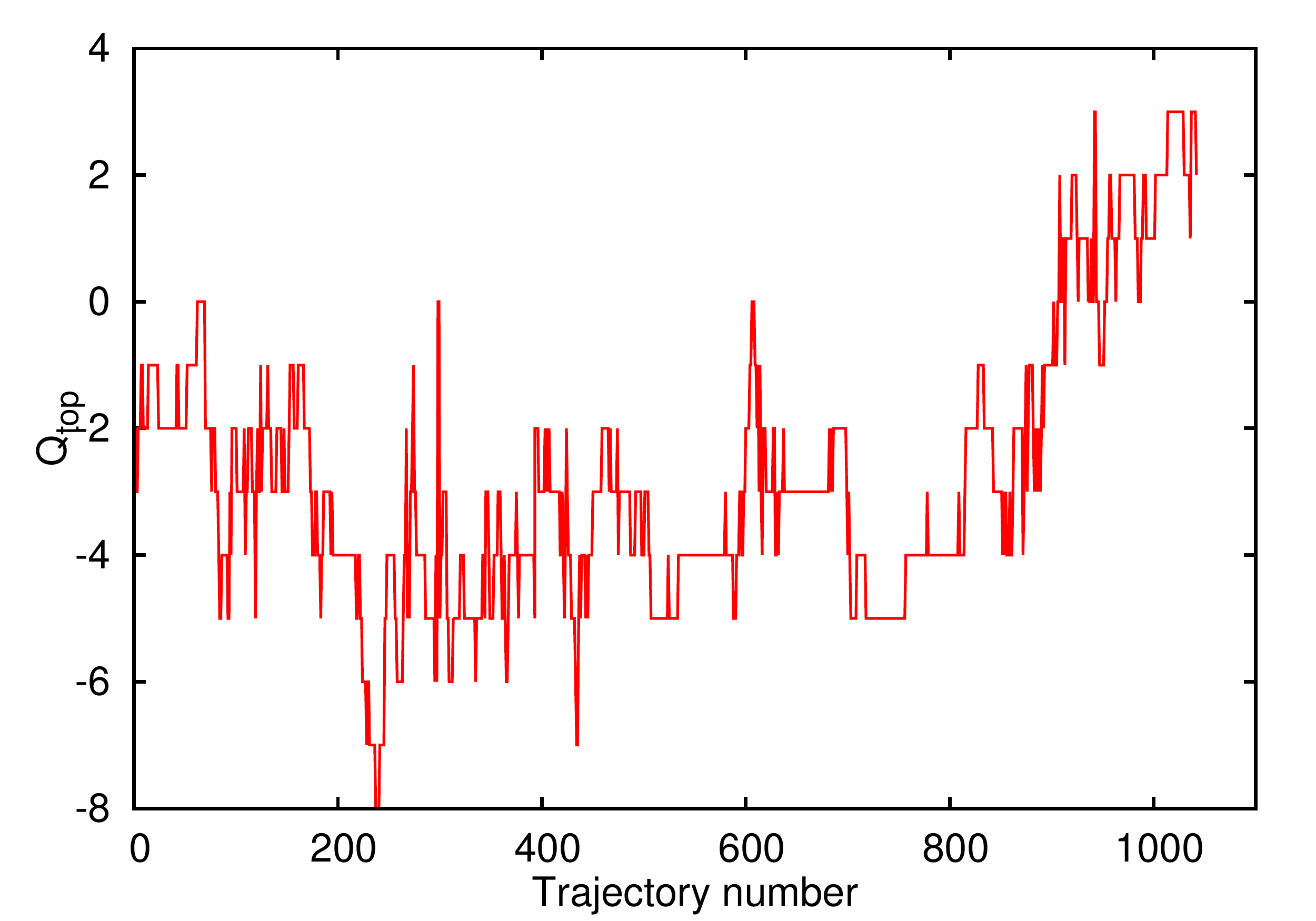}
\caption{The topological index history for one of the $12^324$ dynamical overlap ensembles. }\label{fig:ovchargehist}
\end{figure} 

However, the integrated autocorrelation time for the squared topological charge for this lightest mass ensemble was around 40 trajectories (of length 0.5), compared to around 8 for the plaquette. 
This problem in global topological autocorrelation can be seen in \fig\ref{fig:ovchargehist}. It can also be seen that, although topological index changes occur commonly, there still exist problematic long term correlations for the topological index. Obviously this is an issue for further algorithmic improvements, which, however, should not influence the present analysis significantly. 
 
\section{Results}

\subsection{Visualization of the densities}
A direct way of comparing the topological structures is to plot three-dimensional slices 
of the topological charge density. It is crucial to notice that a direct comparison of 
the profiles from different actions is only possible if the lattice spacings are similar 
because the densities $q(x)$ strongly scale with $a$, namely as $a^{-4}$.

\begin{figure}[t!!]
 \centering
 \setlength{\tabcolsep}{0pt}
\begin{tabular}{ccc}
 \textbf{dynamical overlap} & \textbf{asqtad}& \\
\begin{minipage}[]{0.215\textwidth}
\includegraphics[width=0.9\textwidth]{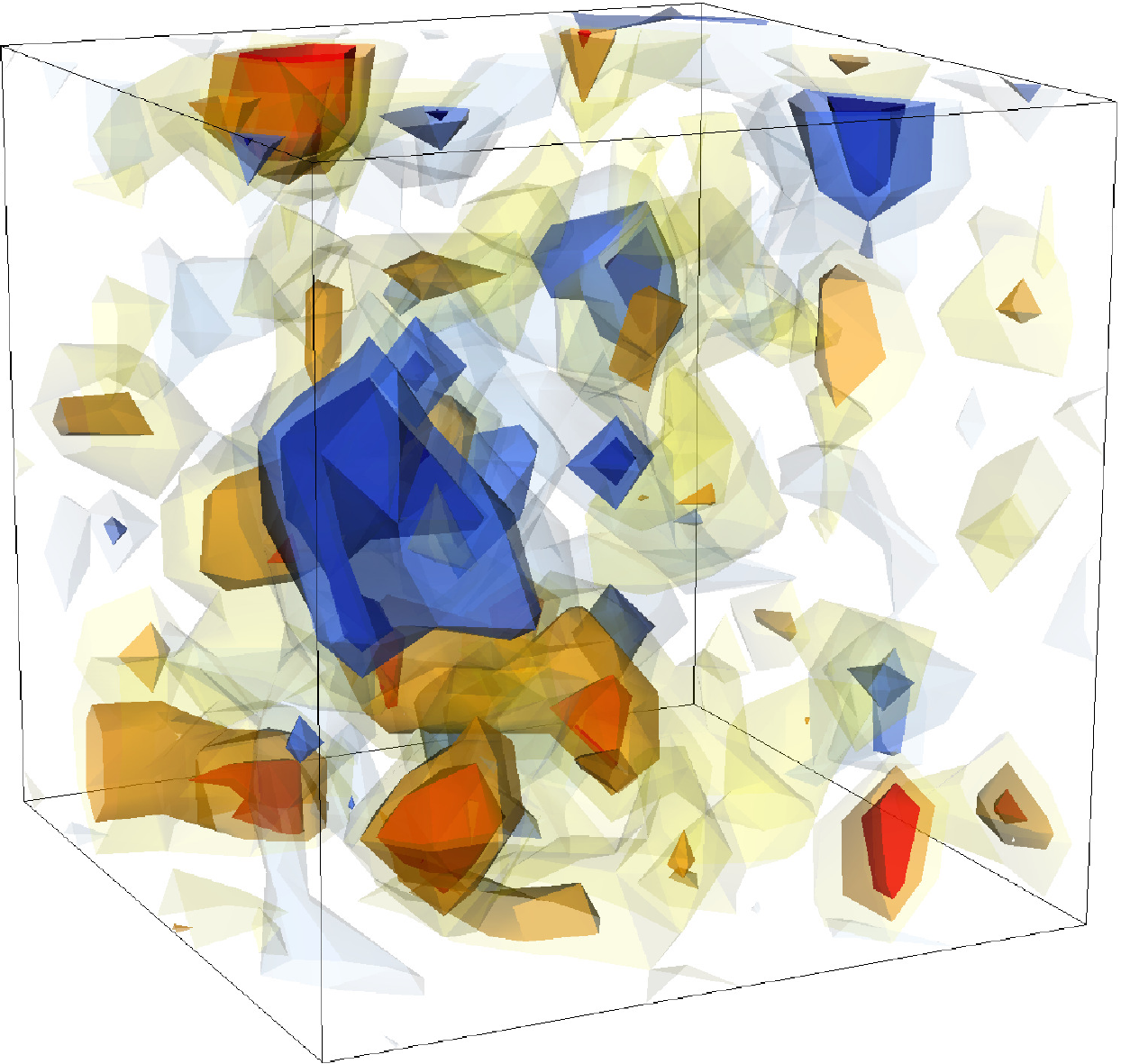}
\end{minipage}  & 
\begin{minipage}[]{0.215\textwidth}
\includegraphics[width=0.9\textwidth]{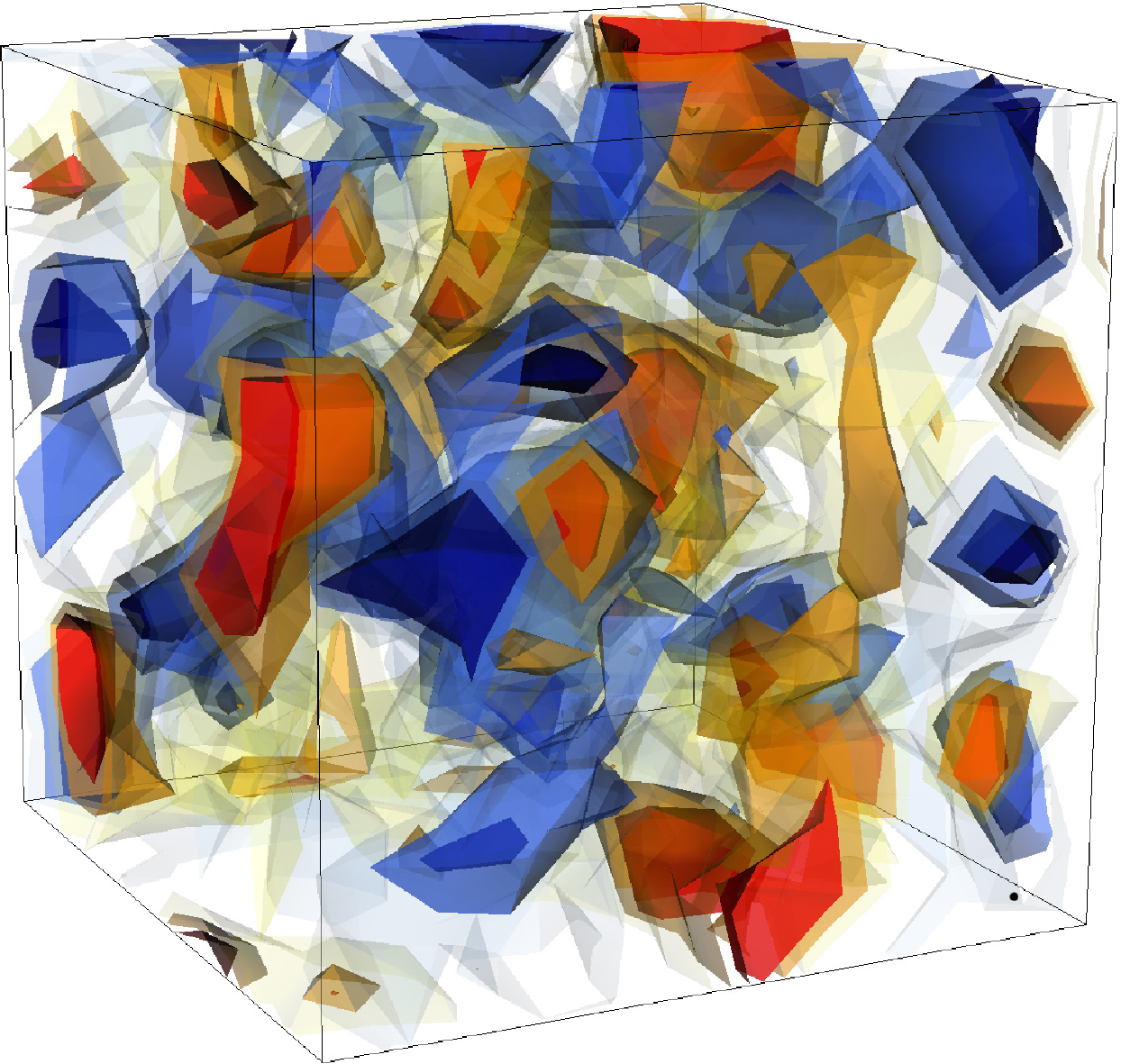}
\end{minipage}&\\[24pt]
\textbf{top. fixed overlap} &\textbf{quenched Iwasaki}&\\
\begin{minipage}[]{0.215\textwidth}
\includegraphics[width=0.9\textwidth]{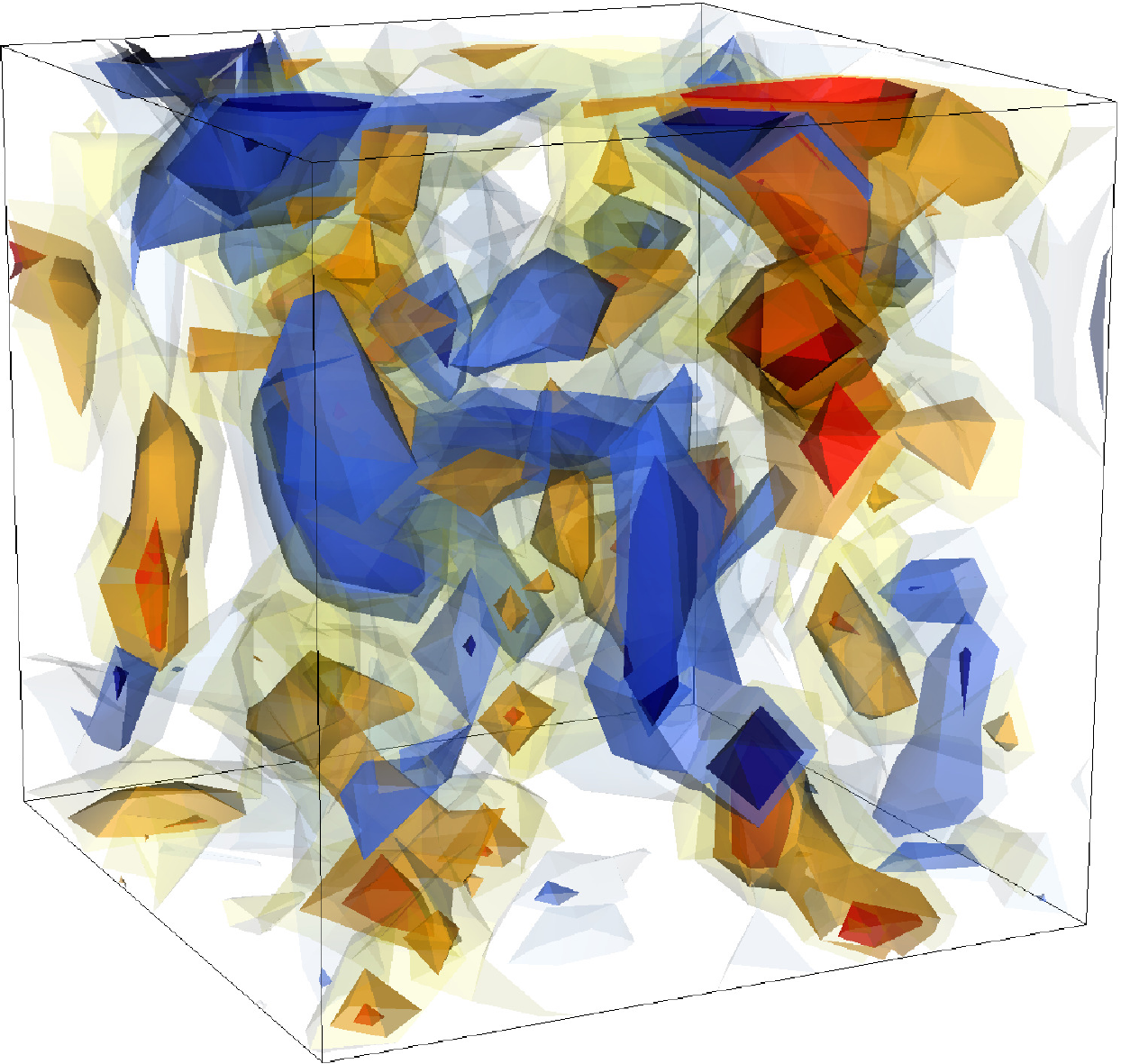}
\end{minipage}  & 
\begin{minipage}[]{0.215\textwidth}
\includegraphics[width=0.9\textwidth]{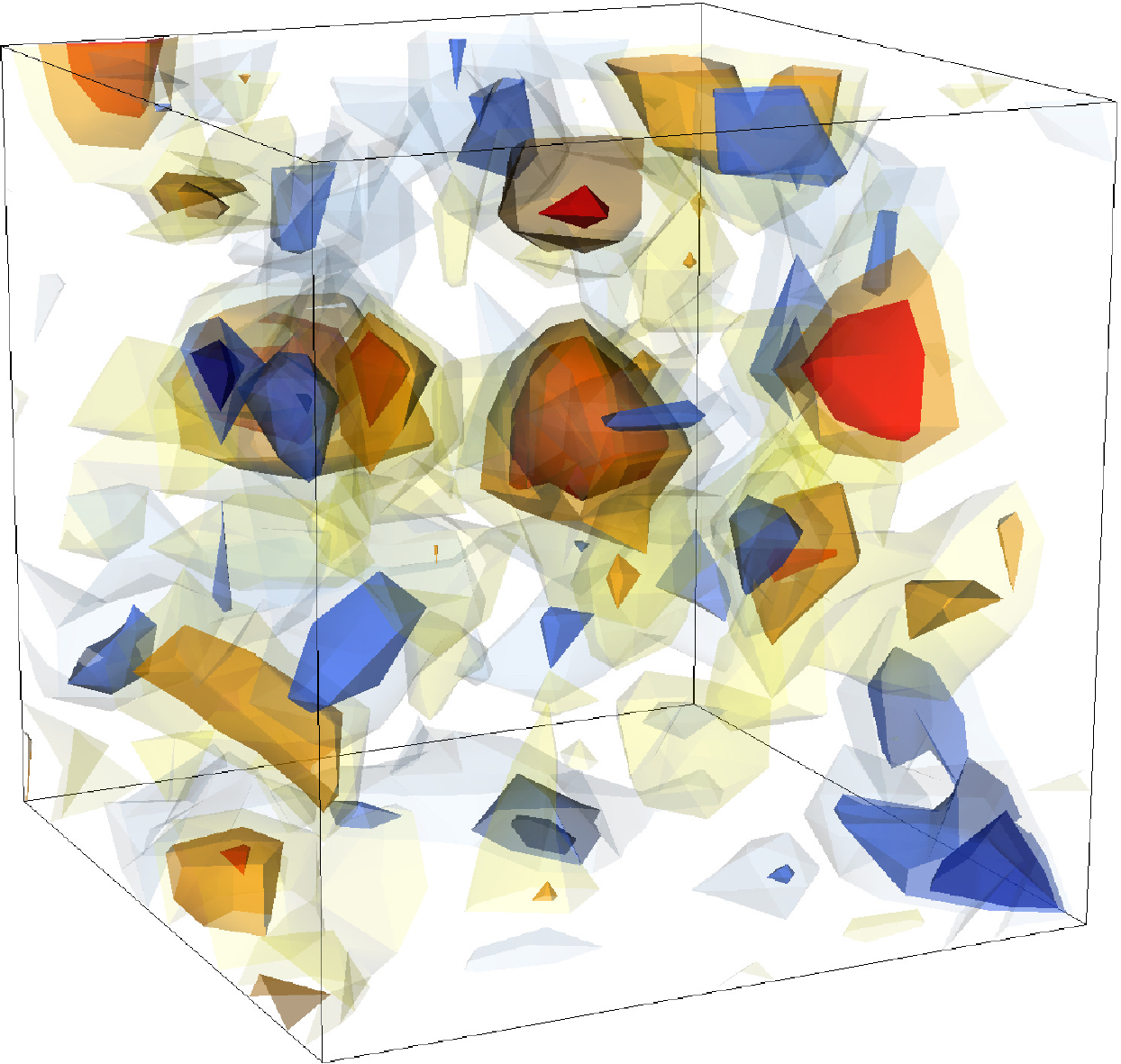}
\end{minipage} &
\\ [24pt]
\textbf{np. clover}& \textbf{quenched plaquette}& \\ 
\begin{minipage}[]{0.215\textwidth}
\includegraphics[width=0.9\textwidth]{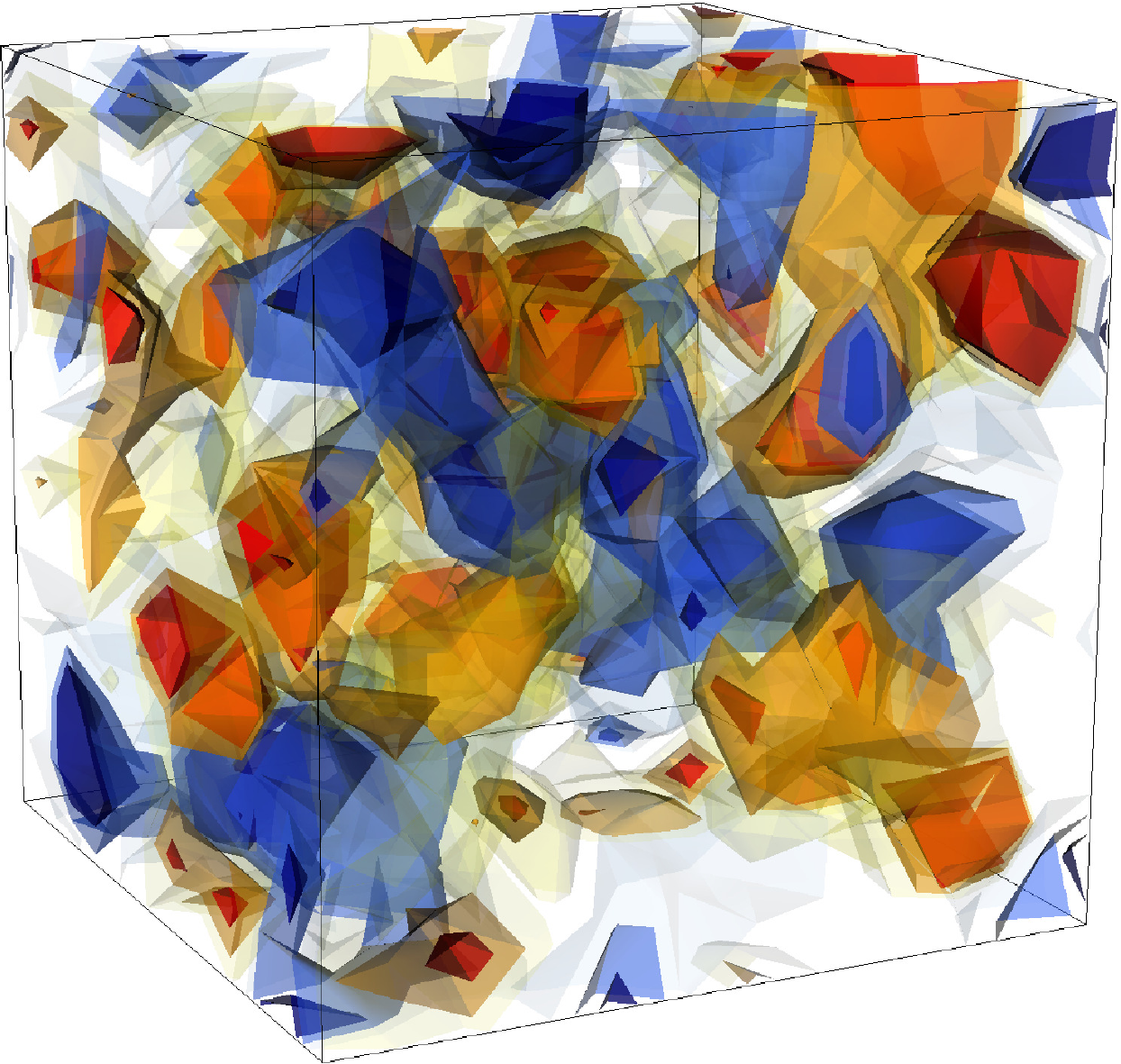}
\end{minipage}& 
\begin{minipage}[]{0.215\textwidth}
\includegraphics[width=0.9\textwidth]{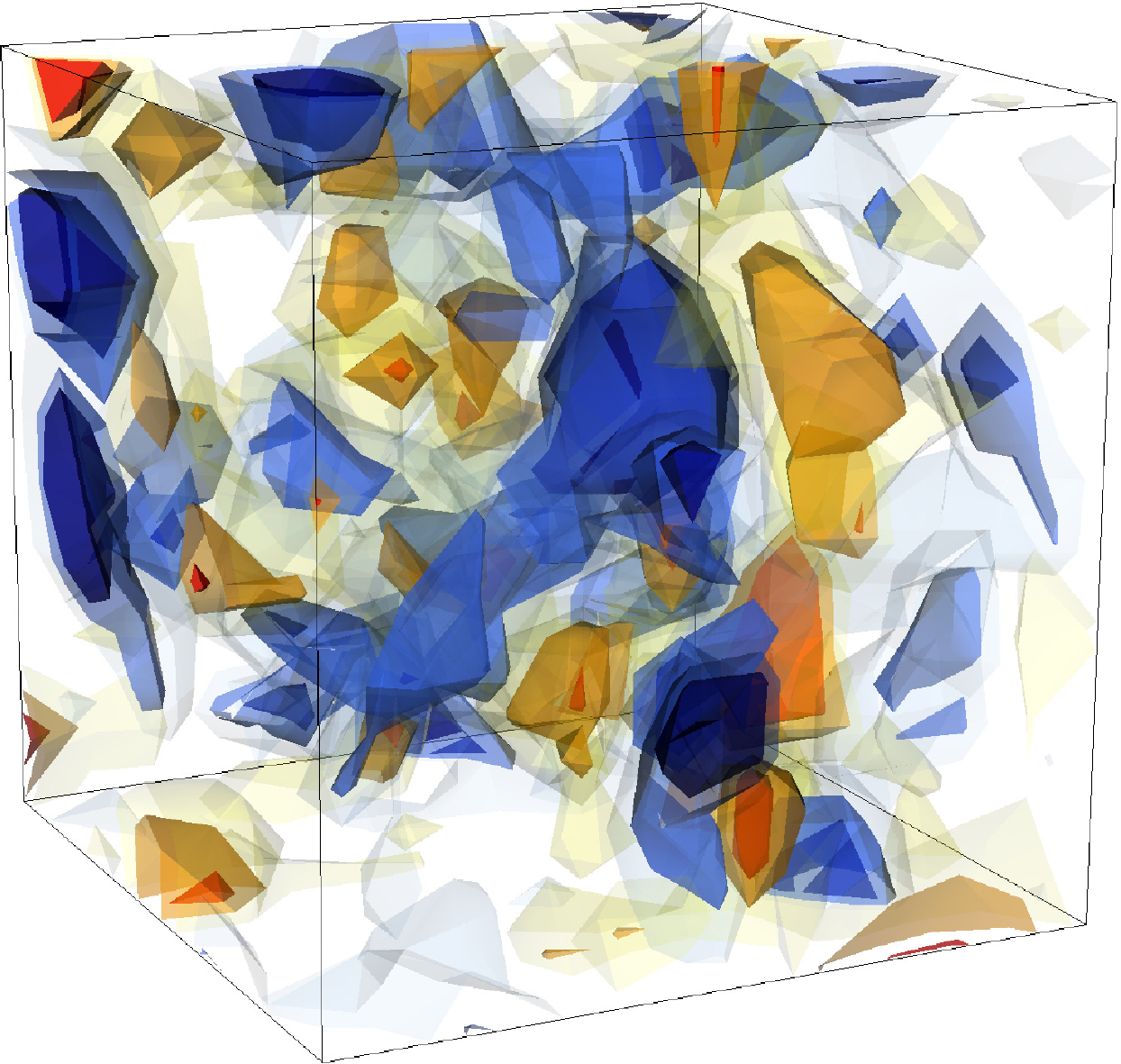}
\end{minipage} & \begin{minipage}[r]{0.05\textwidth}
  \hspace*{-0.1\textwidth}
\includegraphics[width=\textwidth, trim= 20 0 0 0, clip]{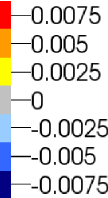}
 \end{minipage}

\end{tabular}
 \caption{Slices of the topological charge for dynamical overlap ($m_{\pi}=600\,\MeV$), asqtad staggered, dynamical overlap with topology fixing term ($m_{\pi}\approx500\,\MeV$ each) and nonperturbative improved clover (quite heavy $m_{\pi}\approx 1\GeV$) fermions, all with the same lattice spacing  $a=0.12 \fm$. The quenched counterparts of the latter two simulation algorithms are also depicted (Iwasaki $\leftrightarrow$ top. fixed overlap and plaquette $\leftrightarrow$ clover). The lattice volume is $12^{3}$ in all cases. This corresponds to a size of approximately $(1.5 \fm)^{3}$ in physical units. The color scale is equal in all plots: Blue represents negative topological charge and red positive charge.} 
 \label{fig:samespacing}
\end{figure}

In \fig\ref{fig:samespacing} we show the topological structure of one sample configuration for different fermion actions, after 5 steps of improved stout smearing. They all have the same lattice spacing 
of $a=0.12\fm$ and the same physical volume $V=(1.44 \fm)^{3}$. 
Below we will discuss how sensitive the topological charge density is to $a$, implying that 
without further analysis one can only compare lattice results for different actions 
if they have the same lattice spacing.  
For comparison we have 
generated quenched systems with two different gauge actions and included their topological 
profiles in that figure. We have chosen the plaquette gauge action and the Iwasaki gauge action, 
which are the quenched counterparts of the nonperturbative clover action and the topology 
fixed overlap configurations, respectively.
While this is only a small fraction of the total four-dimensional volume, one can see 
already some of the main properties analyzed in detail below.
Dynamical lattice simulations tend to give larger fluctuations of the topological charge density 
than quenched ones, as already pointed out in, e.g.,  
Refs.~\cite{Moran2008a,Bruckmann2010a}. This property is not shared by the dynamical overlap results. 
Topology fixed overlap fermions lie somewhat in between. Below we will quantify these observations. 

\begin{figure*}[t!!]
 \centering
\begin{tabular}{cccccc}
\begin{minipage}[]{0.2\textwidth}
\includegraphics[width=1\textwidth]{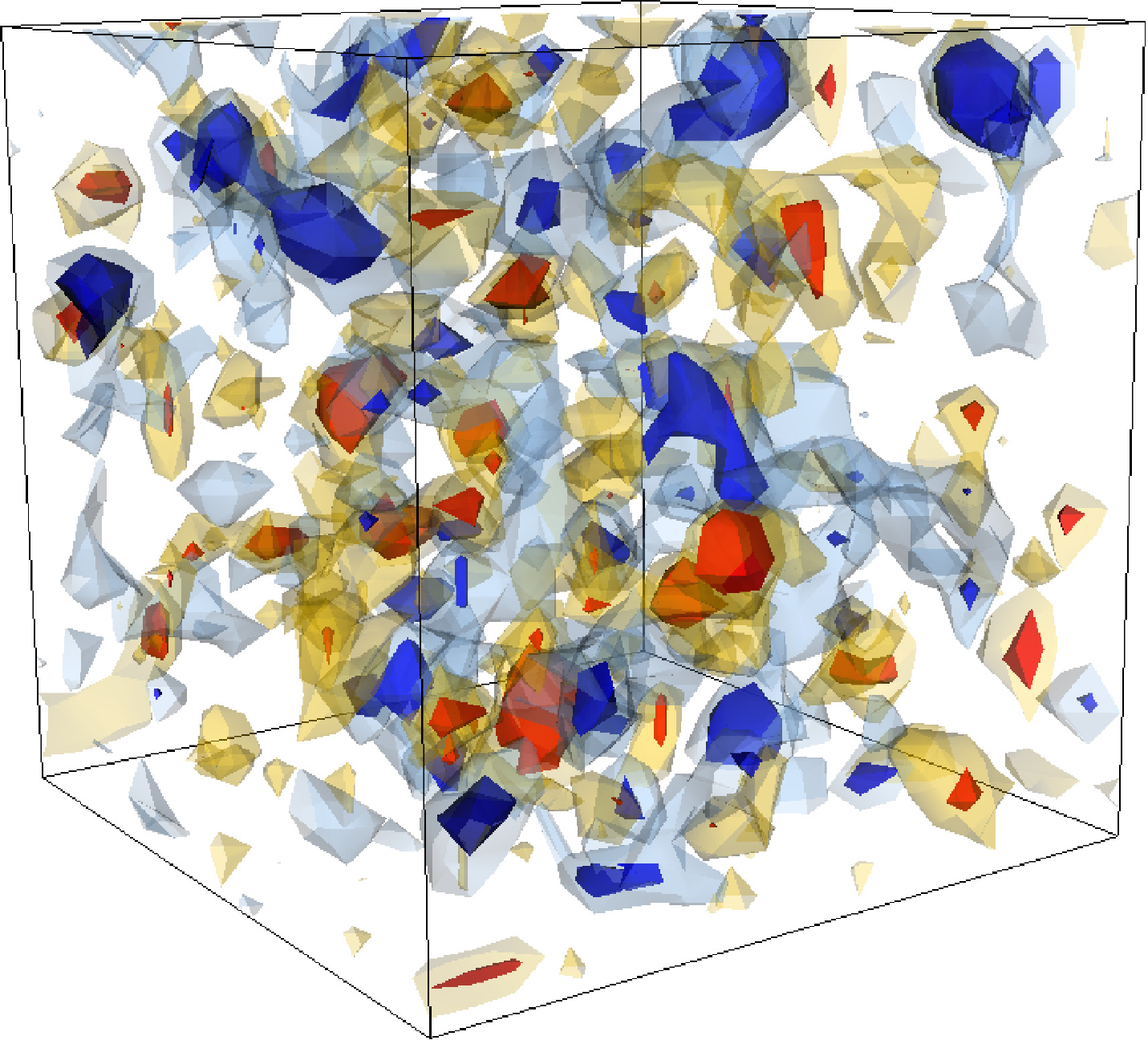}
\end{minipage}  &$\Rightarrow$&\begin{minipage}[]{0.2\textwidth}
\includegraphics[width=1\textwidth]{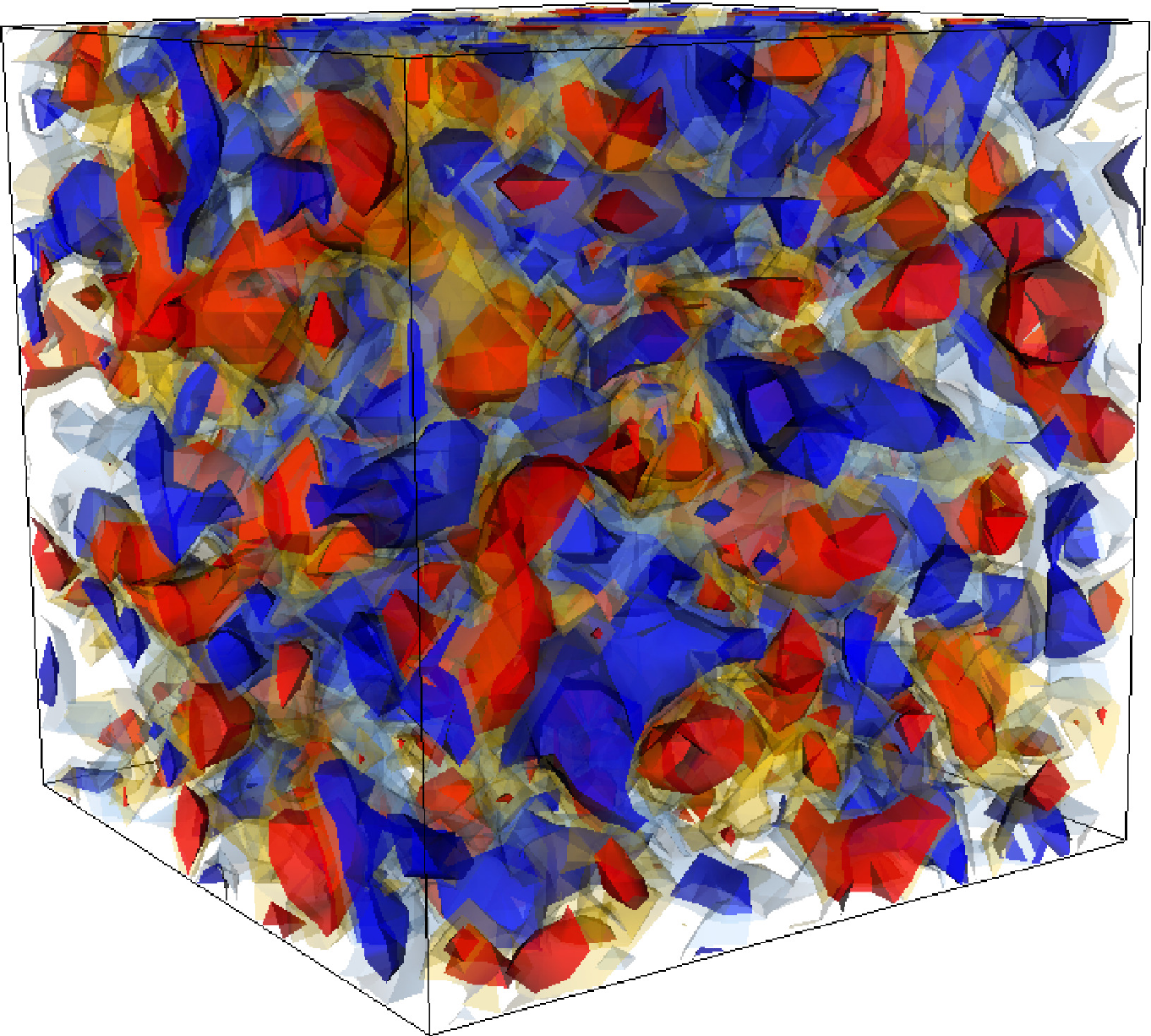}
\end{minipage} &$\Rightarrow$&\begin{minipage}[]{0.2\textwidth}
\includegraphics[width=1\textwidth]{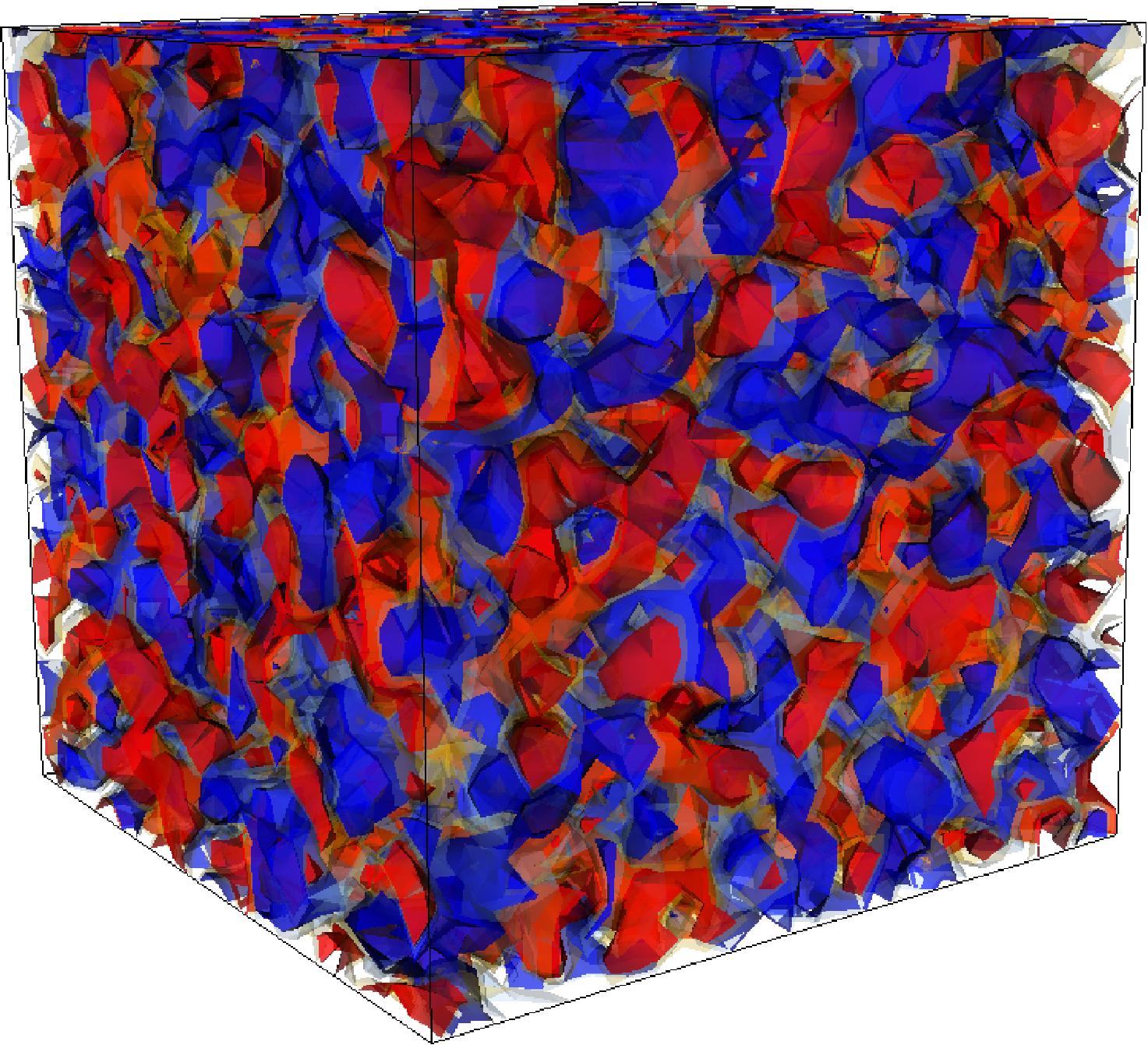}
\end{minipage}& 
\begin{minipage}[r]{0.085\textwidth}
\bigskip
\includegraphics[width=\textwidth]{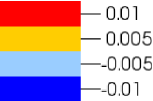}
 \end{minipage} \\ \smallskip\\
\textbf{a=0.1fm}& &\textbf{a=0.079fm} &&\textbf{a=0.063fm} &   
\end{tabular}
 \caption{Three-dimensional slice of the topological charge density for twisted mass fermions at fixed physical volume $V= (1.9 \fm)^{3}$ after 5 steps of improved stout smearing. We see an increasing laminar structure for finer lattices.}
 \label{fig:finerandfiner}
\end{figure*}

To stress the importance of the lattice constant $a$ we show in 
\fig\ref{fig:finerandfiner} the topological charge density for different lattice spacings but fixed physical volume for twisted mass fermions. Two main effects in the continuum limit are clearly visible. First, the structure becomes more and more fine grained and second the magnitude of the density in physical units increases. 

\subsection{Topological clusters}
A cluster analysis of the topological charge density is a powerful tool to characterize the profile of the structure (cf. \cite{Bruckmann2007c,Bruckmann2010a}). A cluster is defined as a set of neighboring lattice points with the same sign of the topological charge density.

To analyze the shape of these clusters, we cut the absolute value of the topological charge density at a variable value $q_{\rm cut}$ and determine the number of clusters $N_{\rm cluster}(q_{\rm cut})$ with $|q(x)|>q_{\rm cut}$ as a function of the total number of points above $q_{\rm cut}$, $N_{\rm points}(q_{\rm cut})$.

It has been shown that this cluster number obeys a power-law and that the exponent
\begin{equation}\label{eq:xi}
\xi= \frac{d \log(N_{\rm cluster}(q_{\rm cut}))}{ d \log (N_{\rm points}(q_{\rm cut}))}
\end{equation}
 of this power-law is highly characteristic for the underlying topological structure. Pure noise, for example, would correspond to $\xi=1$ as every point forms its own cluster and a very smooth density would have a cluster exponent close to zero.
\begin{figure}[t!!]
\centering
\includegraphics[width=\columnwidth]{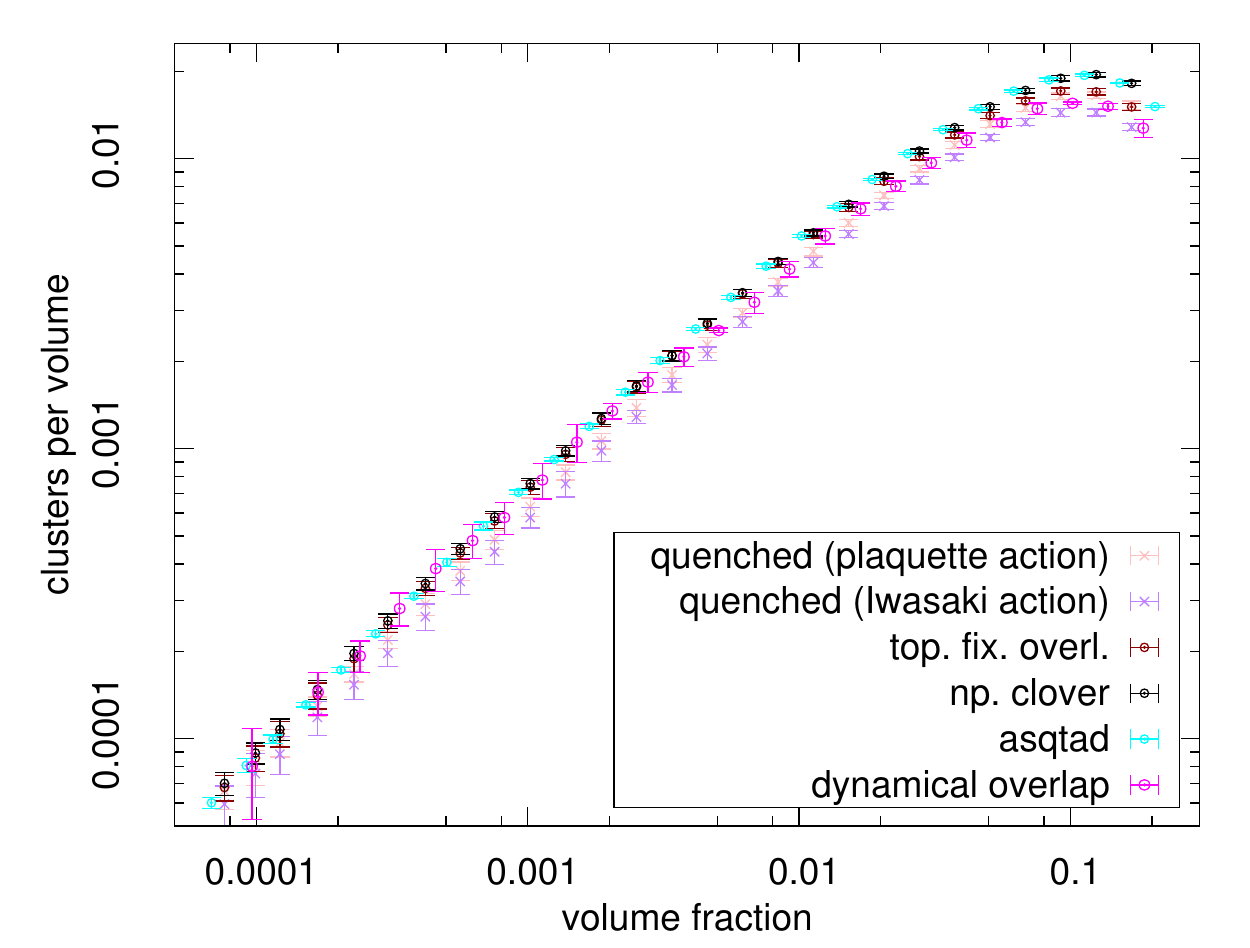}
\caption{The number of clusters per lattice volume as a function of the volume fraction above the cut-off, for the actions used in \fig\ref{fig:samespacing}. Fitting a power-law yields an exponent of $\sim$0.85.}
\label{fig:power-law} 
\end{figure} 

In \fig\ref{fig:power-law} we show this power-law for the actions used in the previous section. We find that the power-laws agree very well and that the exponents are all compatible with each other. Therefore, the topological profiles are similar and the apparent difference in the three-dimensional visualizations in \fig\ref{fig:samespacing} originates from a difference in the absolute values of the densities.

\subsection{Topological charge density correlator} 
\begin{figure}[t!!]
\centering
\includegraphics[width=\columnwidth]{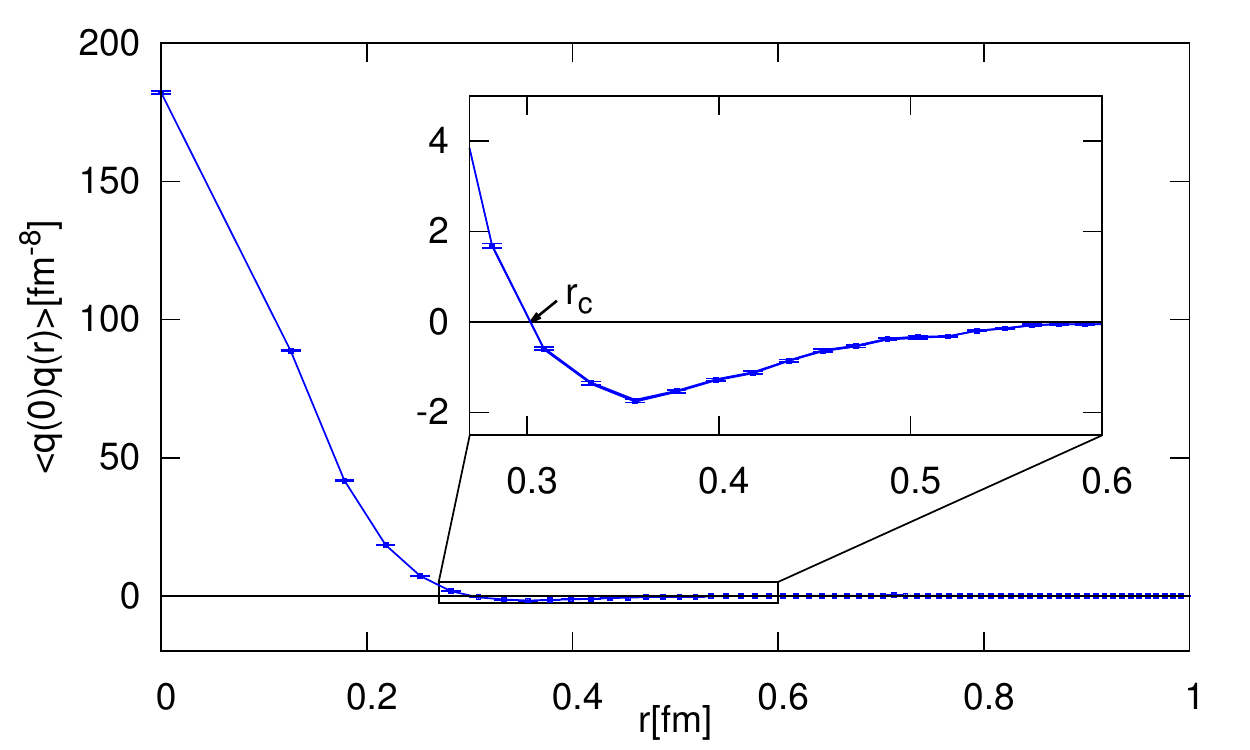}
 \caption{Two-point function of the topological charge density after 5 improved stout smearing steps for an ensemble of dynamical overlap cofigurations.}
 \label{fig:q0qxexample}
\end{figure}
Another possibility to quantify the topological structures is to look at the two-point correlation function of the topological charge density. To this end, we compute the ``all-to-all`` correlator of the density (see \eq(\ref{eq:qdenscorr})) after 5 steps of improved stout smearing. At this point we want to recall that the resulting density is very similar to the fermionic topological charge density \cite{Ilgenfritz:2008ia} and therefore, we find the same qualitative behavior for the two-point function as in Ref.~\cite{Horvath2005}. We show an example of the correlator for the dynamical overlap ensemble in \fig\ref{fig:q0qxexample} 

At small distances the correlator develops a positive core of radius $r_{\rm c}$, for large distance the correlator is compatible with zero and in between it is slightly negative~\cite{Horvath2005}. This behavior is characteristic for all fermion actions and quenched ensembles \cite{Ilgenfritz:2008ia}.

\subsubsection{Size of positive core}

The radius of the positive core $r_c$ is given by the (first) zero of the correlator. To this end, we 
interpolate the correlator between the adjacent data points. 
Although, there are many potential error sources, the position of this zero 
seems to be rather robust and has an excellent signal to noise ratio.
Most importantly $r_c$ shows only little dependence on $a$ when expressed in lattice units,
which allows to compare configurations generated for different lattice spacings. 

 \begin{figure*}[th!]
 \centering
 \includegraphics[width=0.7\textwidth]{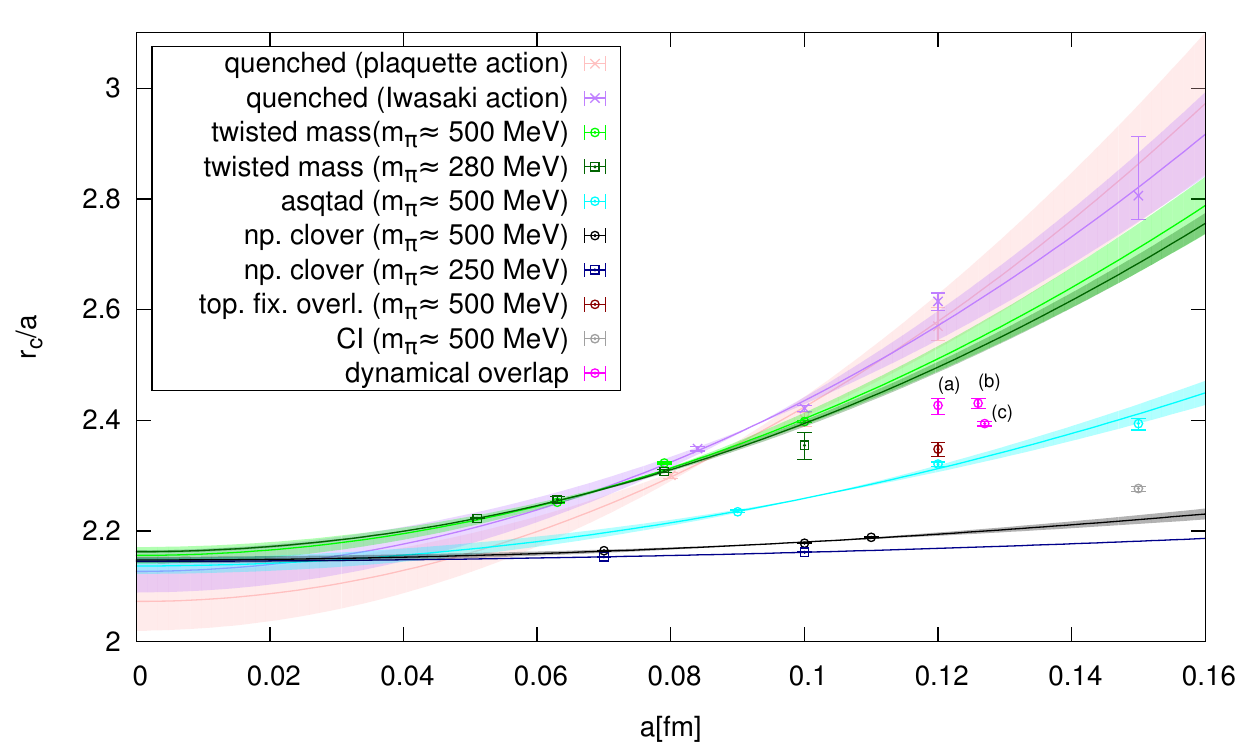}
 \caption{Zero of the 2-point function of the topological charge density. The errors -- which are partly to small to be visible -- result from the interpolation between the points of the correlator. We included fits and error bands for twisted mass, staggered and clover action. There are three different points from the overlap ensembles at different pion masses: (a) $600\,\MeV$, (b) $560\,\MeV$ and (c) $510\,\MeV$. Note that the twisted mass results for $m_{\pi}=280\, \MeV$ and $500\, \MeV$ nearly fall on top of one another.}
 \label{fig:zeroinlatticeunits}
\end{figure*}

We measured this $r_{c}$ for different lattice spacings, while keeping the pion mass approximately constant. 
\fig\ref{fig:zeroinlatticeunits} is our main result from which we will primarily draw our conclusions
on the relevance of chiral symmetry for topological properties.  

\begin{table}[t!!]
\centering
\begin{tabular}{cccc}\hline
\textbf{fermion action }& $m_{\pi} [\MeV]$ &$C$&$B [\fm^{-2}]$\\ \hline
twisted mass   &$\approx 500 $&$2.16(2)$&$25(3)$\\ 
twisted mass &$ \approx 280 $&$2.163(2)$&$23.2(8)$\\   
np imp. clover &$\approx 500$&$2.148(6)$&$3.2(6)$ \\  
np imp. clover  &$\approx 250 $&$2.146$(-)&$1.6$(-)\\  
asqtad staggered  &$\approx 500$&$2.14(1)$&$12(1)$\\  \hline
quenched Iwasaki &-&$2.13(5)$&$35(7)$\\
quenched plaquette &-&$2.07(4)$&$31(4)$\\\hline
\end{tabular}
\caption{Parameters of the single fits. (Only two lattice spacings were available for the light nonperturbative clover ensembles  and, hence, no fit error can be given.)}\label{tab:fitdata}
\end{table}
The core size has to vanish in the continuum limit. We expect discretization/smearing errors which 
are independent of the action used and errors related to these actions. As all actions are $\BigO(a)$-improved
the latter should only set in at  $\BigO(a^2)$. Therefore, we fit the data with a function of the form 
\begin{equation}
    r_{c}/a=C+B\cdot a^{2}\,.
\end{equation}
We find indeed that the coefficients $C$ are for all actions compatible with $C=2.15$ and that the $B$'s differ, see \tab\ref{tab:fitdata}. The size of these differences is what we interested in.
We did not fit higher order terms in the universal part, because our data points were not sufficient to do so
and because we are only interested in the differences between the actions.
We expected that actions with better chiral properties should give results which are markedly closer to the dynamical 
overlap ones than actions with strong violation of chiral symmetry.  

Let us further discuss the universality of this scaling behavior. First of all,
if we compare the curves for twisted mass and clover fermions at $m_{\pi}\approx 500\, \MeV$ and  $m_{\pi}\approx 250\,\MeV$, 
we find a relatively weak dependence on the mass.

Secondly, we have found the constant $C$ to depend on smearing (not shown). This is not surprising as the positive part of the 
topological charge correlation gets smeared out over more lattice spacings.

\fig\ref{fig:zeroinlatticeunits} also includes the results for dynamical overlap fermions with and without topology fixing term 
at similar lattice spacings. The data points for dynamical overlap configurations are not on one line, because they belong to 
different pion masses and, hence, different fit curves. 
 
We have also included two quenched results for comparison. The extrapolations yield a consistent value for the constant $C$ with a slightly larger error than the dynamical configurations. The quenched results  are in accordance with the results of Horvath \etal~\cite{Horvath2005}. They used the full fermionic 
definition of the topological charge density (not truncated in Dirac modes) and find a core size $r_{c}\approx 2a$ 
for the Iwasaki gauge action. 

The non-universal term $Ba^2$ measures how $r_c/a$ from different actions converges towards $C$. 
The spread at current lattice spacings $a\leq 0.15\fm$ is on the order of $10\%$. The overlap 
data lie in the middle of that range and even the quenched results are not far off. Our conclusion 
is that the different actions with their different treatment of chiral symmetry do not differ much 
with respect to this important topological observable. Consequently, all of them can be used 
to study topological quantities. To the extent that topological properties are relevant for 
hadron properties this observation also ties in with the fact that quite often the differences between 
quenched and dynamical results for these are not huge (typically of the order 10-30\%). 

\subsubsection{Contact term}  

 \begin{figure}[t]
\centering
 \includegraphics[width=\columnwidth]{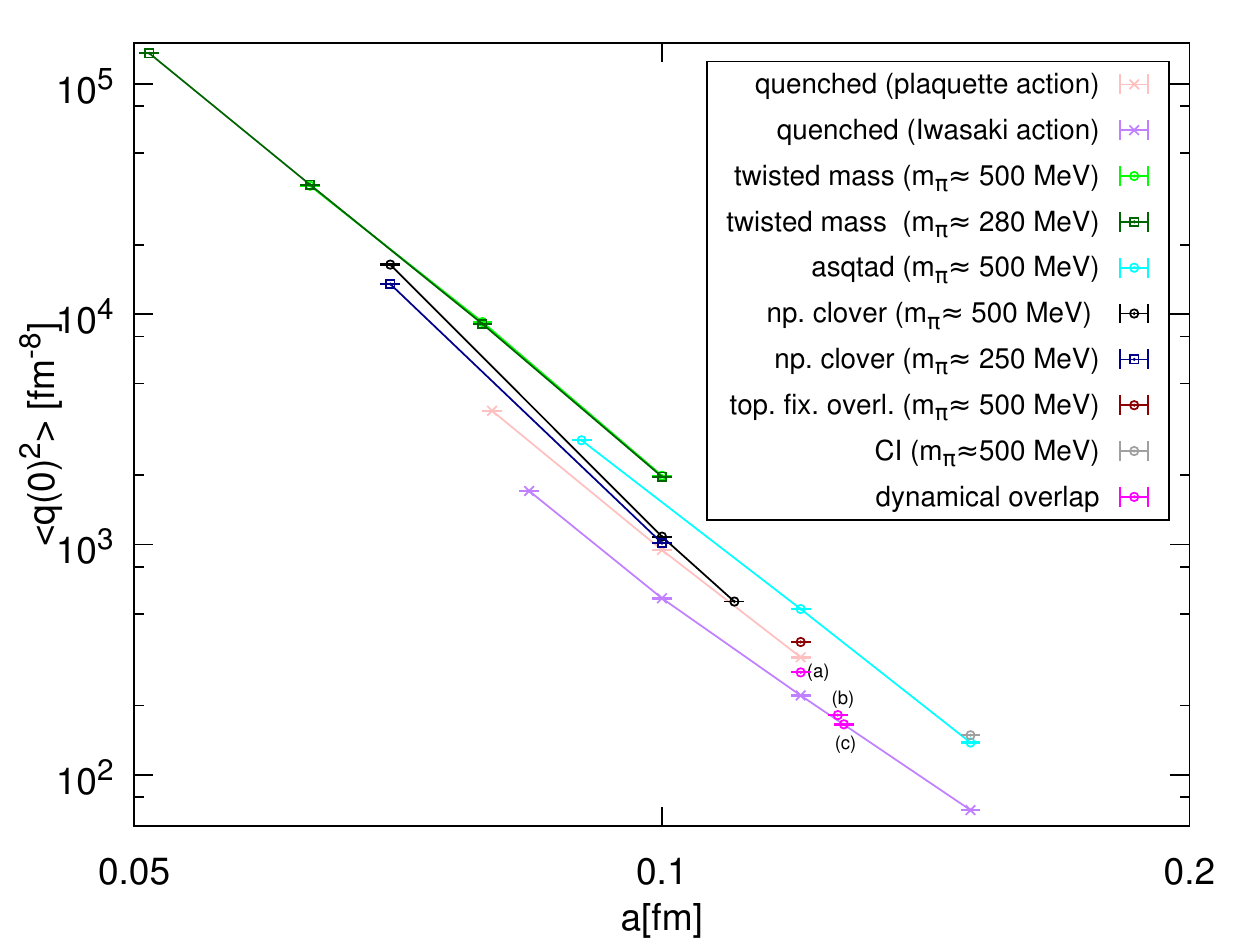}\\
 \caption{Maxima of the two point function of the topological charge density. Dynamical overlap results are labeled by $m_{\pi}$ as: 
(a) $600\,\MeV$, (b) $560\,\MeV$ and (c) $510\,\MeV$. 
Note that the twisted mass results for $m_{\pi}=280\,\MeV$ and $500\,\MeV$ fall on top of one another.}
 \label{fig:maxima} 
\end{figure}

As explained in the introduction, the two-point function has to develop a positive contact term in the continuum limit. 
Therefore, the mean-square value of the topological charge density $\langle q(0)^{2}\rangle$ has to be divergent 
in the continuum limit. \fig\ref{fig:maxima} shows a double logarithmic plot of $\langle q(0)^{2}\rangle$ 
in physical units versus the lattice spacing. The linear behavior indicates a power-like divergence for $a\to0$ for all actions:
\begin{equation}
 \langle q(0)^{2}\rangle \propto a^{-c}
\end{equation}
for some positive number $c$.

This exponent is similar for the different actions and its value is around $-6$. Only the nonperturbative clover action deviates from this value with an exponent around $-7$. As the contact term is highly divergent in the continuum limit, we do not expect the fitted values to agree and, therefore, we do not present them.
 
The contact term of the dynamical ensembles is bigger than for their quenched counterparts, but it also important to notice, that the contact term for dynamical overlap configurations at $a=0.12 \fm$ lies between the two quenched simulations (plaquette and Iwa\-saki). Thus we can conclude, that dynamical fermions generate indeed larger fluctuations 
in the vacuum at finite lattice spacing, as argued in \cite{Moran2008a}, but also that the differences between different actions are as large as those between quenched and dynamical simulations.

\section{Summary} 

We have investigated the topological charge density for state-of-the-art lattice actions with dynamical fermions, 
including new dynamical overlap simulations. This quantity was chosen because of the intimate connection between 
topology and chiral symmetry. Different fermion actions do generate different topological landscapes, as visualized 
in \fig\ref{fig:samespacing} and quantified through the topological charge correlator. The change in the 
topological observables is not very large. The radius of the positive core of the topological 
correlator $r_c$ approaches zero with the same slope $C$ for all actions. 
In the next-to-leading order $r_c$ differs, but the spread 
is below $10\%$ and even quenched simulations do not produce markedly different results.
In particular, simulations with exact overlap fermions give results which are 
quite similar to those obtained with topology fixed overlap fermions. The differences between quenched and dynamical 
simulations are not larger than those between different dynamical fermion actions. 
Also, the topological charge density 
seems to be little affected by changes in pion mass.  In contrast, the effects for $\langle q(0)^{2}\rangle$ 
are large but unsystematic. These results are very sensitive to the lattice spacing $a$,
implying that one should be very careful not to jump to conclusions when comparing topological properties of different 
configurations. If we use our dynamical overlap results as benchmark for the quality of the other actions with respect to chirality we 
have to conclude that all of them are reasonable successful and none of them seems to be clearly superior. The differences between results for dynamical overlap fermions and topology 
fixed overlap fermions are especially small, as one might have expected. 

 We wish to acknowledge the International Lattice Data Grid (ILDG) for providing most of the configurations. We would like to thank Gunnar Bali and Ernst-Michael Ilgenfritz for insightful discussions and Johannes Najjar for helping to download the configurations from the ILDG. Furthermore, we want to thank C.B. Lang, M. Limmer and D. Mohler who generated the chirally improved configurations which were used in this Letter. F.B. has been supported by DFG BR 2872/4-2 and F.G. by SFB TR-55 as well as by a grant of the ``Bayerische Elitef\"orderung``. We thank NIC, J\"ulich and LRZ, Garching for providing the computer time used to generate the dynamical overlap fermion and CI configurations, respectively. This research was supported in parts by the European Union under Grant Agreement number 238353 (ITN STRONGnet).


\bibliographystyle{elsarticle-num}
\bibliography{overlaptopology}
\end{document}